\documentclass[twocolumn]{aastex631}

\usepackage[flushleft]{threeparttable}
\usepackage{booktabs}
\usepackage{amsmath}

\begin{document}

\title{DDO~68-C: HST confirms yet another companion of the {\it isolated} dwarf galaxy DDO~68}

\author[0000-0001-6464-3257]{Matteo Correnti}
\affiliation{INAF Osservatorio Astronomico di Roma, Via Frascati 33, 00078, Monteporzio Catone, Rome, Italy}
\affiliation{ASI-Space Science Data Center, Via del Politecnico, I-00133, Rome, Italy}
\email{matteo.correnti@inaf.it}

\author[0000-0003-3758-4516]{Francesca Annibali}
\affiliation{INAF - Osservatorio di Astrofisica e Scienza dello Spazio di Bologna, via Piero Gobetti 93/3, 40129 Bologna, Italy}

\author[0000-0001-8200-810X]{Michele Bellazzini}
\affiliation{INAF - Osservatorio di Astrofisica e Scienza dello Spazio di Bologna, via Piero Gobetti 93/3, 40129 Bologna, Italy}

\author[0000-0002-4775-7292]{Mariarosa Marinelli}
\affiliation{Space Telescope Science Institute, 3700 San Martin Drive, Baltimore, MD 21218, USA}

\author[0000-0003-4137-882X]{Alessandra Aloisi}
\affiliation{Space Telescope Science Institute, 3700 San Martin Drive, Baltimore, MD 21218, USA}
\affiliation{Astrophysics Division, Science Mission Directorate, NASA Headquarters, 300 E Street SW, Washington, DC 20546, USA}

\author[0000-0001-6291-6813]{Michele Cignoni}
\affiliation{Dipartimento di Fisica, Università di Pisa, Largo Bruno Pontecorvo 3, 56127, Pisa, Italy}
\affiliation{INAF - Osservatorio di Astrofisica e Scienza dello Spazio di Bologna, via Piero Gobetti 93/3, 40129 Bologna, Italy}
\affiliation{INFN, Largo B. Pontecorvo 3, 56127, Pisa, Italy}

\author[0000-0002-0986-4759]{Monica Tosi}
\affiliation{INAF - Osservatorio di Astrofisica e Scienza dello Spazio di Bologna, via Piero Gobetti 93/3, 40129 Bologna, Italy}

\author[0000-0002-6389-6268]{Raffaele Pascale}
\affiliation{INAF - Osservatorio di Astrofisica e Scienza dello Spazio di Bologna, via Piero Gobetti 93/3, 40129 Bologna, Italy}

\author[0000-0002-1821-7019]{John M. Cannon}
\affiliation{Macalester College, 1600 Grand Avenue, Saint Paul, MN 55105, USA}

\author[0009-0007-3163-3678]{Lila Schisgal}
\affiliation{Macalester College, 1600 Grand Avenue, Saint Paul, MN 55105, USA}

\author[0000-0001-9162-2371]{Leslie K.~Hunt}
\affiliation{INAF-Osservatorio Astrofisico di Arcetri, Largo E. Fermi 5, 50125, Firenze, Italy}

\author[0000-0001-5618-0109]{Elena Sacchi}
\affiliation{Leibniz-Institut für Astrophysik Potsdam (AIP), An der Sternwarte 16, 14482 Potsdam, Germany}

\author[0000-0001-8368-0221]{Sangmo Tony Sohn}
\affiliation{Space Telescope Science Institute, 3700 San Martin Drive, Baltimore, MD 21218, USA}
\affiliation{Department of Astronomy \& Space Science, Kyung Hee University, 1732 Deogyeong-daero, Yongin-si, Gyeonggi-do 17104, Republic of Korea}




\begin{abstract}
We present the results of deep Hubble Space Telescope photometry of the dwarf galaxy DDO~68-C, proposed as possibly associated with the isolated peculiar dwarf DDO~68. The new data resolve for the first time the stars of DDO~68-C down to well below the tip of the Red Giant Branch (RGB), revealing a low mass (M$_{\star}\simeq 1.5\times 10^7~$M$_{\sun}$) star forming galaxy with a backbone of old stars. By means of a fully homogeneous analysis and using the RGB tip as a standard candle, we find that DDO~68 and DDO~68-C lie at the same distance from us, within the uncertainties ($D=12.6\pm 0.3$~Mpc and $D=12.7\pm 0.4$~Mpc, respectively), thus confirming that the two dwarfs are physically associated. While paired dwarf galaxies with mutual projected distance similar to DDO~68 and DDO~68-C are not exceptional in the Lynx-Cancer Void where they live, DDO~68 remains a unicum as, in addition to the newly confirmed companion, it records the evidence of at least two other satellites.
\end{abstract}




\section{Introduction} 
\label{sec:intro}

Extremely metal-poor dwarf galaxies (XMPs) in the nearby Universe, with metallicity as low as a few percent Solar \citep{Guseva2015} and with active star formation, are of crucial importance for a better understanding of the Universe evolution, because they represent the only opportunity to study the details of star formation and chemical evolution in a regime similar to that of primordial galaxies in the early Universe. Some low-mass XMPs can be explained with the combined effects of inefficient star formation and high metal loss via galactic winds triggered by supernova explosions in shallow potential wells \citep[][and references therein]{Skillman2013,Hirschauer2016,Annibali2022}, but relatively massive XMPs are quite difficult to understand and challenge models of galaxy evolution \citep{Izotov2018,McQuinn2020}. These systems are strong outliers in the luminosity-metallicity (L-Z) and mass-metallicity (M-Z) relations, with a measured metallicity far too low for their stellar mass/luminosity. Accretion of metal-poor gas from the intergalactic medium and/or from interaction with smaller companions are among the mechanisms proposed to explain their anomalous low metal content \citep{Ekta2010,McQuinn2020,Annibali2022}.

A very intriguing case of XMP with metallicity strongly deviating from the M-Z relation is the dwarf irregular galaxy DDO~68, located at a distance of $D\simeq13$ Mpc \citep{Tikhonov2014,Sacchi2016}, whose H~II regions’ oxygen abundance of just $\simeq3\%$ Solar \citep{Pustilnik2005,Annibali2019b} appears incompatible with its stellar mass of M$_{\star} \simeq 10^8 $M$_{\odot}$ \citep{Sacchi2016}. DDO~68 is located in a huge Void \citep[the Lynx-Cancer Void,][]{Pustilnik2011}, but it has been shown to have merged with a ten-times smaller gas-rich companion (hereafter DDO~68-B or the B component), responsible for its very distorted morphology characterized by a large “cometary tail” \citep{Tikhonov2014,Sacchi2016}. Moreover, a second, even smaller (M$_{\star} \simeq 10^6$ M$_{\odot}$) interacting satellite (hereafter S1), likely a gas-free spheroidal galaxy \citep[][hereafter A19]{Annibali2019a} has been discovered thanks to deep photometry acquired with both the Large Binocular Telescope (LBT) and, subsequently, with the Hubble Space Telescope (HST). DDO~68 is therefore the smallest dwarf galaxy with clear evidence for accretion of at least two satellites. 

Nevertheless, this complex merging history seems unable to fully explain DDO~68’s extremely low oxygen abundance: detailed N-body and hydro-dynamical simulations show that, even assuming that the measured metallicity in DDO~68’s H~II regions is associated with the companion’s accreted material still inefficiently mixed with that of the host galaxy, chemical abundances as low as the observed ones are difficult to explain \citep{Pascale2022}.
Understanding the origin of the extremely low metallicity in DDO~68, and in XMPs in general, is fundamental for our understanding of galaxy evolution: XMPs are being discovered with increasing success over the past few years, and nearly all of them are outliers in the M-Z and L-Z relations \citep{Skillman2013,Hirschauer2016,Annibali2019a,Hsyu2017,Yang2017,Senchyna2019,McQuinn2020,Pustilnik2021}. 

\begin{figure*}[thbp!]
\begin{center}
{\includegraphics[width=0.9\textwidth]{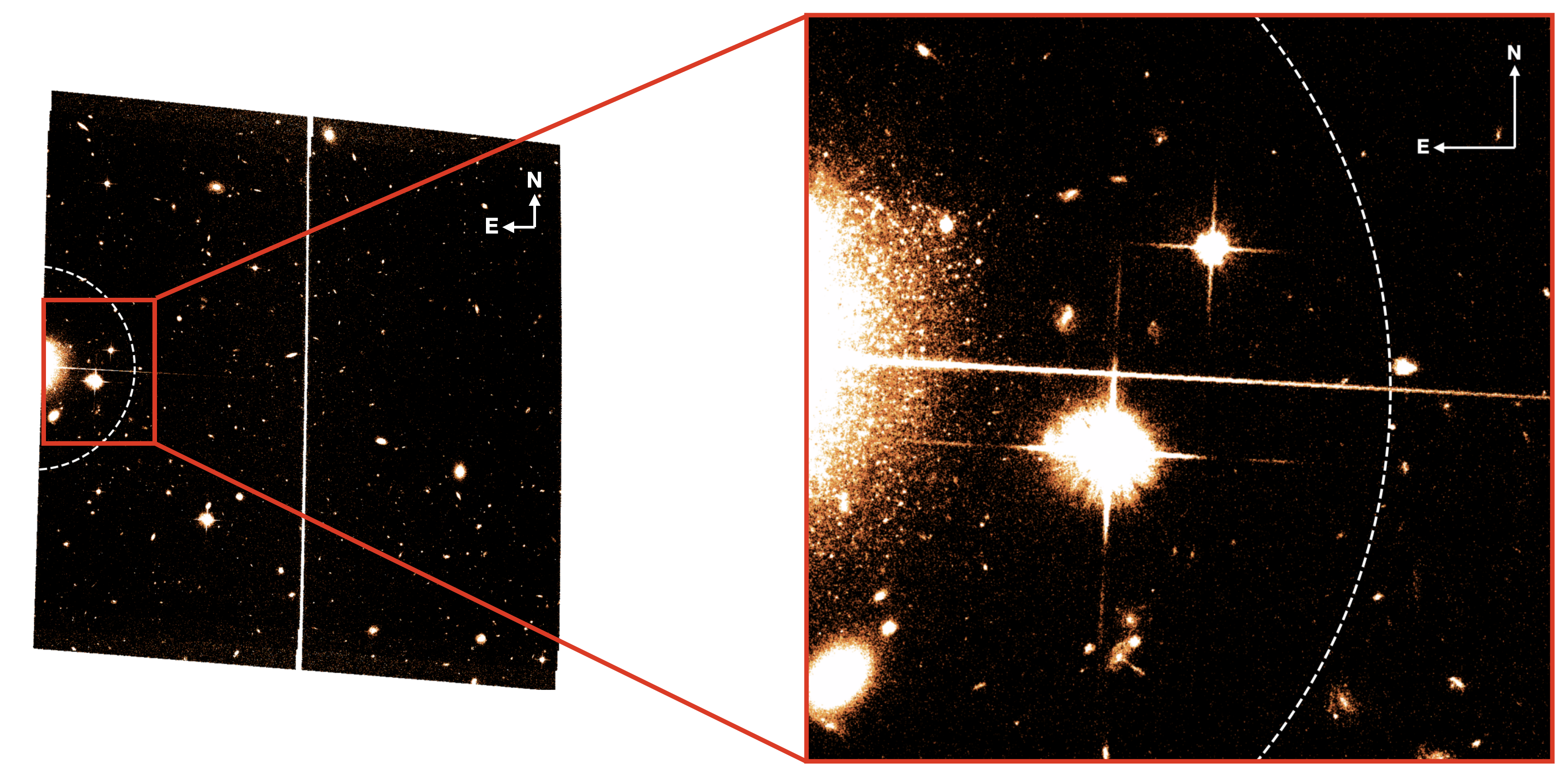}}
\end{center}
\caption{Left: F814W ACS/WFC field of view. DDO~68-C is located at the edge of the WFC chip, in order to avoid as much as possible the strong contamination from TYC 1967-1114-1. The white dashed semicircle (r = 800 px, corresponding to $\sim$ 2.5 kpc) indicates the region where we consider sources in the final catalog as belonging to DDO~68-C. Right: zoom-in of the ACS/WFC field of view in the DDO~68-C region, showing how the galaxy stars are clearly resolved in the ACS/WFC image. The diffraction spike from TYC 1967-1114-1 is clearly visible, cutting the stellar body of DDO 68-C approximately in half.}
\label{fig:fov} 
\end{figure*}

DDO~68 is one of the handful of XMP outliers in the M-Z relation that is close enough to allow for a detailed study of its evolutionary history based on a resolved star approach.
A possible solution to the problem of its low metallicity was proposed by \cite{Cannon2014}, when they discovered a candidate third interacting system with Very Large Array (VLA) 21-cm observations. Their detected gas-rich (M$_{HI} \simeq 2.8\times 10^7$M$_{\odot}$) system, dubbed DDO~68-C, is at only 11$^{\prime}$ from DDO~68, has its same systemic velocity, and appears connected to DDO~68 by a low surface brightness H~I bridge. This led the authors to suggest that if the two galaxies are located at about the same distance, since they are only 43~kpc apart, in projection, they may have interacted in the past. On these grounds, \cite{Cannon2014} suggested that an infall of low metallicity gas into DDO~68 from a close encounter with DDO~68-C could have provided sufficient gas dilution to explain the deviation of DDO~68 from the M-Z relationship. Assuming the same distance as the main body, the gas mass of DDO~68-C equals $\simeq 3\%$ of the gas mass in DDO~68, but it may have been much higher in the past before being deposited into its more massive companion during the interaction. Unfortunately, nothing is known so far about the stellar or dynamical masses of DDO~68-C, making its physical properties very poorly constrained. Most importantly, any plausible scenario invoking an interaction with DDO~68 requires a robust determination of DDO~68-C’s distance. 

Until now, the distance of DDO~68-C could not be derived on safe grounds, mostly because the object was not detected in optical bands, due to the presence of a bright foreground red star along its line of sight \citep[TYC 1967-1114-1, with spectral type K0V–K2V and a magnitude of K = 8.5;][]{Cannon2014,Annibali2023}. Fortunately, GALEX FUV and NUV images, where the foreground red star is not hiding the candidate system as much as in redder bands, allowed \cite{Cannon2014} to detect diffuse emission co-located with the H~I object, thus revealing the presence of rather recent star formation. This is, however, insufficient to provide useful information on stars older than a few hundred Myr or on the galaxy’s distance. Hence, the hide-and-seek situation of DDO~68-C makes a study of its resolved stellar content impossible from the ground in seeing-limited mode. 

A preliminary way out of the impasse was proposed by exploiting the high spatial resolution offered by adaptive optics. \cite{Annibali2023} acquired deep imaging of the galaxy in the J and H bands with the LUCI-SOUL instrument on the LBT, and showed that bright individual stars are well resolved in the adaptive optics images and are indeed mostly concentrated in correspondence with the peak of the FUV emission. The H versus J-H color-magnitude diagram (CMD) inferred from those data exhibits a sparse distribution of stars but does not reach the magnitude depth of the typical discontinuity associated with the RGB tip (hereafter TRGB) detection. In fact, under the assumption of a 13 Mpc distance, the LBT photometry of the stars resolves blue loop and Asymptotic Giant Branch (AGB) stars with ages $\simeq$50 Myr to $\simeq$1.5 Gyr old, but remains at least 1 mag brighter than the TRGB expected magnitude \citep{Annibali2023}.

While the low surface brightness H~I gas that connects DDO~68 and DDO~68-C offers tantalizing evidence for an ongoing interaction \citep{Cannon2014}, only HST observations could place the physical association of the two systems on a more secure footing. We then acquired deep HST imaging of DDO~68-C, and we present here the results of its photometric study, allowing us to resolve its individual stars and determine its distance through the TRGB. With DDO~68-C at the same distance as DDO~68, the system is a unique case of a small dwarf interacting with three satellites: really {\it a flea with smaller fleas that on him prey}\footnote{From Jonathan Swift$'$s {\it On Poetry: a Rhapsody}: So, naturalists observe, a flea/ has smaller fleas that on him prey;/ and these have smaller still to bite ‘em/ and so proceed ad infinitum.}, as \cite{Annibali2016} dubbed it. Multiple interactions with an unexpected large population of satellites may also provide a viable explanation to the anomalous extremely low metallicity observed in DDO~68, as well as in many other XMPs.

In Section~\ref{sec:reduction} we describe our new HST observations and the corresponding data reduction. In Section~\ref{sec:cmd} we present the CMD of DDO~68-C and describe its stellar content. In Section~\ref{sec:trgb} we derive the galaxy distance through the analysis of the TRGB, whereas in Section~\ref{sec:structure} we estimate its structural parameters. Finally, in Section~\ref{sec:discussion} we discuss the results obtained in this work, 
which are briefly summarized in Section~\ref{sec:summary}.

\section{HST photometry of DDO~68-C} 
\label{sec:reduction}

DDO~68-C was imaged with the Wide Field Channel (WFC) of the Advanced Camera for Surveys (ACS) in Cycle 30 (GO-17131, PI: F. Annibali) in the F606W and F814W bands. As demonstrated by previous studies \citep{Tully13}, this is the best filter combination to efficiently detect the TRGB in galaxies a few Mpc away, and allows the study of both the young and the old stellar populations. 

To avoid its effects, observations were planned such that TYC 1967-1114-1 was placed $\simeq 5$ arcsec outside the edge of the ACS/WFC detector, paying attention to avoid the regions where the Dragon’s breath
\citep{Porterfield2016} and scattered light effects are predicted to be the highest, while keeping the bulk of DDO~68-C within the camera field of view (see left panel of Figure~\ref{fig:fov}). This configuration allowed us to mitigate these effects, but could not prevent a higher background level due to the wings of the star PSF and the presence of the star's spikes, clearly visible in the targeted field. The brightest spike cuts DDO~68-C into half, as can be seen in the right panel of Figure~\ref{fig:fov}, which shows the field of view zoomed on DDO~68-C. However, with careful selections in the photometric catalogs, as described below, we were able to remove its contribution in the final CMD. 

\begin{figure*}[thbp!]
\centering
{\includegraphics[width=0.45\textwidth]{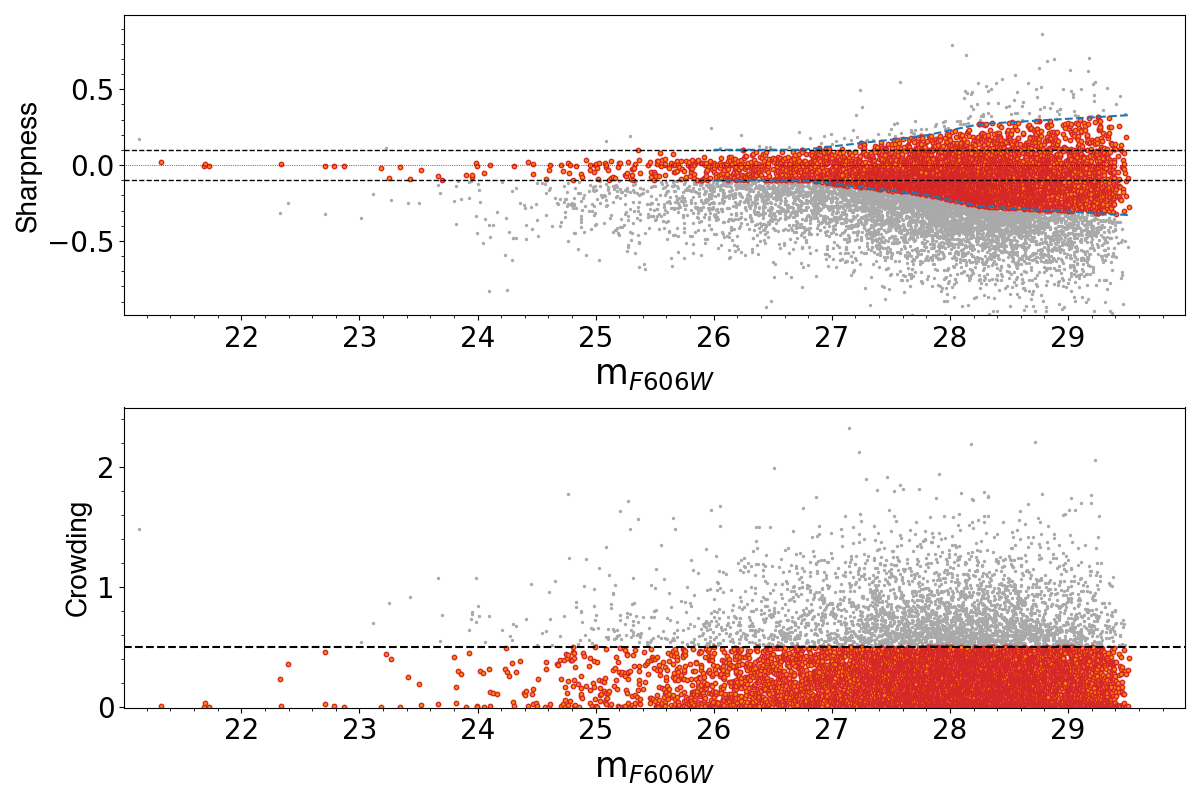}}
{\includegraphics[width=0.45\textwidth]{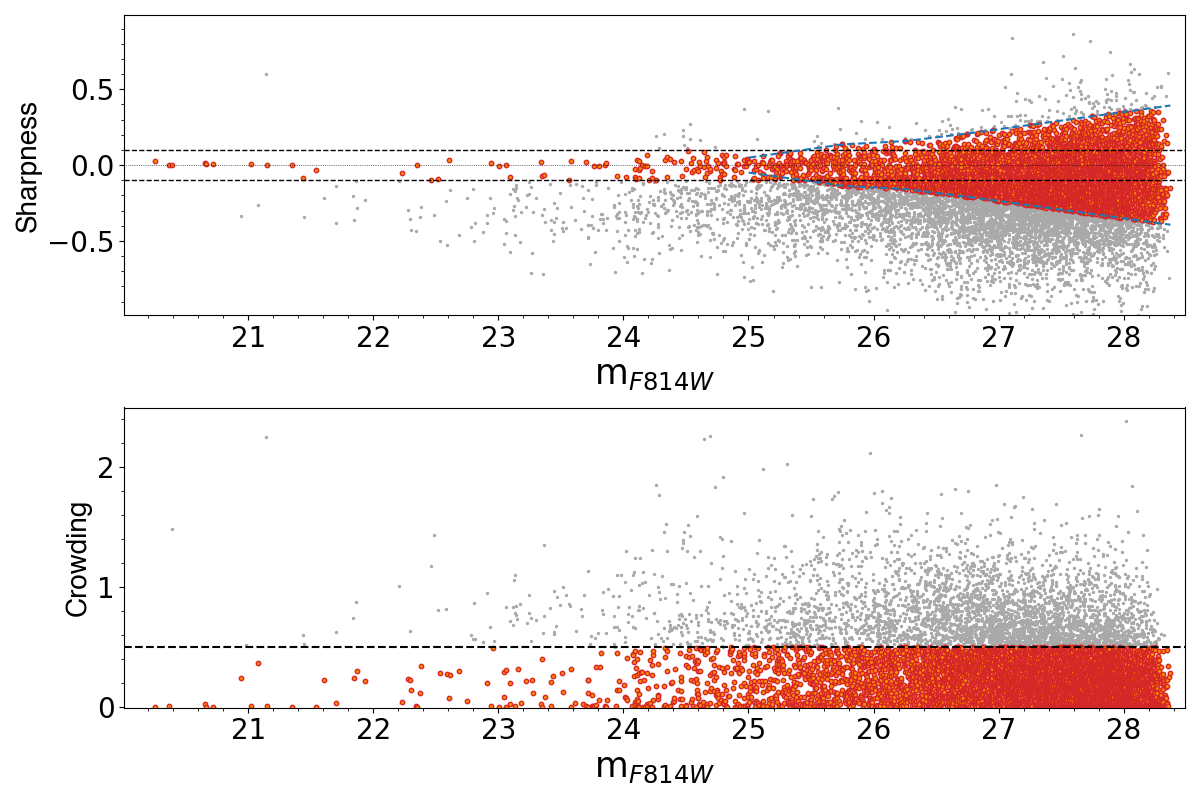}}
\caption{Dolphot {\tt shaepness} (top panels) and {\tt crowding} (bottom panels) as a function of m$_{F606W}$ and m$_{F814W}$ magnitudes (left and right panels, respectively). Light gray dots represent all sources that pass the preliminary selection cuts, as described in Section~\ref{sec:reduction}, while red points depict the sources that pass our {\tt sharpness} and {\tt crowding} cuts, reported as dashed lines.}
\label{fig:sel_stars} 
\end{figure*}

To perform stellar photometry, we adopted the same approach as in A19 and references therein. Briefly, we first downloaded from the HST archive\footnote{\url{https://mast.stsci.edu/portal/Mashup/Clients/Mast/Portal.html}} the {\it flc} science images, which correspond to the bias-corrected, dark-subtracted, flat-fielded, CTE corrected, GAIA-aligned images. We combined them into a single-stacked, distortion-corrected image (to be used as a reference frame in the photometric data reduction). Then, we used the latest version of DOLPHOT \citep[][and references therein]{Dolphin2000, Dolphin2016}  to obtain simultaneous multi-filter PSF photometry, calibrated in the Vegamag system. We set the DOLPHOT parameters using a hybrid combination between the default values and those derived by \cite{Williams2014}. 

To exclude remaining artifacts and spurious detections from the DOLPHOT output, we adopted selection cuts based on the diagnostic parameters included in the photometric catalog. In particular, after an initial selection where we retained only sources with flag {\tt Object type} $\leq$ 1, {\tt Photometry quality flag} $\leq$ 2, and {\tt SNR} $>$ 3, we adopted additional cuts in {\tt sharpness} (tracing the difference of the source size with respect to point sources) and {\tt crowding} (tracing the impact of light contamination from nearby sources on the flux measurement) that are effective in removing a larger fraction of contaminants and badly measured stars. Here, the adopted selection in {\tt sharpness} deserves particular attention: in fact,  due to the presence of the prominent foreground star's spike, as well as of other bright saturated sources, our photometric catalog contains a larger fraction of objects with negative {\tt sharpness} values compared to typical photometry outputs (i.e., nearly symmetric bell-shaped distribution around {\tt sharpness}  = 0.0), preventing us to adopt typical selections as those described in \cite{Annibali2019b}. 

We thus adopted the following alternative approach: in the {\tt sharpness}  versus magnitude distribution, we selected objects with positive {\tt sharpness}, divided them in 0.5 mag bins, and derived the 90$^{th}$ percentile from the cumulative distribution in each of them. We then interpolated the discrete values and retained the sources below the curves or with {\tt sharpness} $<$ 0.1. We then mirrored the selection for the negative {\tt sharpness}. This selection is illustrated in the top panels of Figure~\ref{fig:sel_stars} (m$_{F606W}$, left panel, m$_{F814W}$, right panel). As for {\tt crowding}, a straight cut at  $<$ 0.5 was adopted, as shown in the bottom panels of Figure~\ref{fig:sel_stars}.

In order to remove residual spurious detections associated with the star spike and sources not belonging to DDO~68-C, we further applied a position cut, selecting only stars that are located in a semicircle centred at the position (X, Y) = (2180, 40) and with a radius of 800 px, corresponding to $\sim$ 2.5~kpc, at the distance of  DDO~68, depicted as the white dashed line in Figure~\ref{fig:fov}. After this last selection, our final catalog contains 1259 sources. 

In order to ensure that our comparison with DDO 68 is as homogeneous as possible, we reduced the HST-ACS images described and analyzed by \cite{Sacchi2016} exactly in the same way and adopting the same selections as described above for DDO~68-C. The final DDO~68 catalogue contains 72111 sources. 

\begin{figure*}[thbp!]
\begin{center}
{\includegraphics[width=0.95\textwidth]{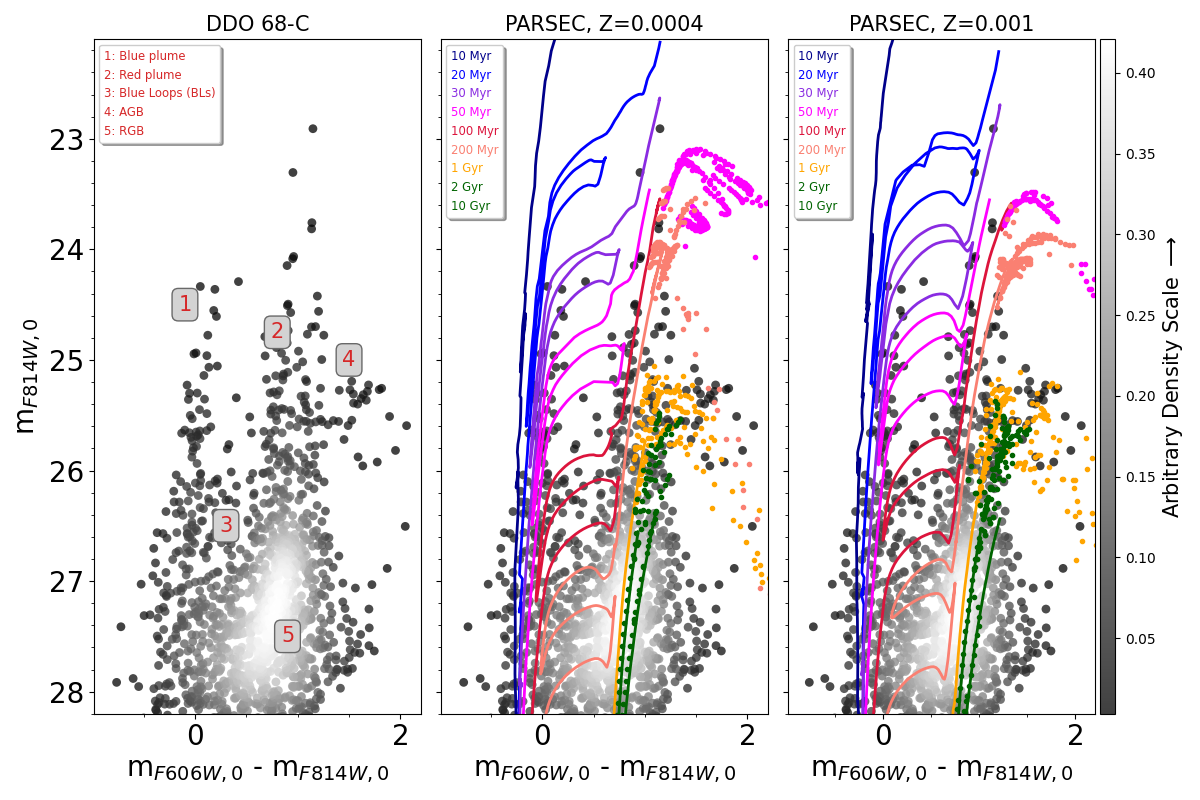}}
\end{center}
\caption{Reddening-corrected m$_{F814W,0}$ versus m$_{F606W,0}$-m$_{F814W,0}$ CMD of DDO~68-C, with points color-coded according to the star count density in the CMD. The different evolutionary phases, described in Sect.~\ref{sec:cmd}, are indicated in the left panel. The PARSEC stellar isochrones \citep{Bressan2012,Marigo2017} in the ACS WFC bands are overplotted on the CMD for two different metallicities of Z=0.0004, or $\sim$2\% Solar (middle panel), and Z=0.001, or $\sim$6\% Solar (right panel), and for stellar ages in the range 10\,Myr$-$10\,Gyr, as indicated in the legend. The isochrones were shifted adopting a distance modulus of $(m-M)_0$=30.52 mag. For sake of clarity, dotted lines are used to display, for isochrone ages  $\leq$ 1 Gyr, the TP-AGB phase and, for isochrone ages $\geq$2 Gyr, the AGB phase and later ones.}
\label{fig:cmd} 
\end{figure*}

\section{Stellar Content: DDO~68-C CMD}
\label{sec:cmd}

In Fig.~\ref{fig:cmd}, left panel, we show the m$_{F814W,0}$ versus m$_{F606W,0}$ - m$_{F814W,0}$ CMD of DDO~68-C  obtained from the final clean photometric catalog described in Sect.~\ref{sec:reduction}. The plotted CMD was corrected for foreground reddening adopting an E(B$-V$) = 0.016, from the \citet{sfd98} maps re-calibrated according to \citet{sf11}, and using the reddening laws reported in \citet[hereafter BP24]{Bellazzini2024}. Stellar isochrones \citep{Bressan2012,Marigo2017} for two different metallicities of Z=0.0004, ($\sim$2\% Solar) and Z=0.001 ($\sim$6\% Solar), and for stellar ages in the range 10\,Myr$-$10\,Gyr, have been displayed on top of the observed CMD in the central and right panel, respectively. 

The CMD of DDO~68-C exhibits all the stellar evolutionary phases typical of star forming systems: a ``blue plume'', at $-0.4 \lesssim $m$_{F606W,0}$-m$_{F814W,0}\lesssim 0.2$, populated by young ($\lesssim$ 20 Myr) main sequence (MS) stars and evolved post-MS stars in their hot core He-burning phase, aka blue loop (BL); a ``red plume'', at $0.6 \lesssim$ m$_{F606W,0}-$m$_{F814W,0}\lesssim 1.3$, m$_{F814W,0} \lesssim 26.5$, populated by a mix of core-He burning stars at the red edge of the BL and AGB stars, with ages from $\sim$20 Myr to several Gyrs; at intermediate colors, our CMD samples BL stars as old as $\sim$200 Myr, while older BLs fall below the photometric depth achieved; the few bright and very red objects at m$_{F606W,0}-$m$_{F814W,0}\gtrsim 1.3$, m$_{F814W,0}\lesssim 26$ are thermally pulsing asymptotic giant branch (TP-AGB) and Carbon stars with ages from $\sim$300 Myr to $\sim$2 Gyr; last, the most densely populated feature in the CMD, at $0.5 \lesssim$ m$_{F606W,0}-$m$_{F814W,0}\lesssim 1.2$, m$_{F814W,0}\gtrsim 26.5$, is the RGB of low mass stars, corresponding to a population at least as old as $\sim$2 Gyr, and potentially as old as 10-13 Gyr. In Section~\ref{sec:trgb}, the tip of the RGB will be used to infer the distance of DDO~68-C. 

An interesting feature in the CMD of Fig.~\ref{fig:cmd} is the presence of a prominent concentration of stars at  $0.7 \lesssim$ m$_{F606W,0}-$m$_{F814W,0}\lesssim 1.2$, $25.8 \lesssim$ m$_{F814W,0}\lesssim 26.1$, producing a sort of ``discontinuity'' in the red plume, a less prominent concentration of stars in the blue plume at $-0.2\lesssim$ m$_{F606W,0}-$m$_{F814W,0}\lesssim 0.2$, $26\lesssim$ m$_{F814W,0}\lesssim 26.8$, and an ``overdensity'' of intermediate-color stars which seems to connect the two. From a comparison with stellar models, these features appear to be well matched by the BL phase of a  $\sim$100 Myr old stellar population, suggesting a specific event of enhanced activity in the star formation history (SFH) of  DDO~68-C around that epoch.

It is worth highlighting that the presence of a burst of SF in this galaxy may be interesting in light of its possible interaction history with DDO~68, as suggested by previous studies \citep[e.g.,][]{Cannon2014}.

In Fig.~\ref{fig:cmd}, central panel, the adopted stellar metallicity of Z=0.0004 for the displayed models is compatible with the oxygen abundance measured in DDO~68's H~II regions \citep{Pustilnik2005,Annibali2019b} which, as already discussed in Section~\ref{sec:intro}, falls much below the mass-metallicity relation for dwarf galaxies, and thus it has been suggested to be due to the accretion of metal-poor gas from the inter-galactic medium or from very metal-poor systems. 

This metallicity seems to reasonably match the colors of the RGB, whereas the Z=0.001 models predict RGBs that are too red. On the other hand, both metallicities seem to provide a satisfactory fit for ages $\leq$ 1 Gyr, while a metallicity as high as Z=0.004 would result in overly red Red Super Giant stars. This suggests that the oldest populations have metallicity of approximately Z=0.0004, while younger ones have Z$\lesssim$0.004, indicating a modest metal enrichment over time, as is typically observed in dwarf irregular galaxies.

\begin{figure*}[thbp!]
\center{{\includegraphics[width=0.45\textwidth]{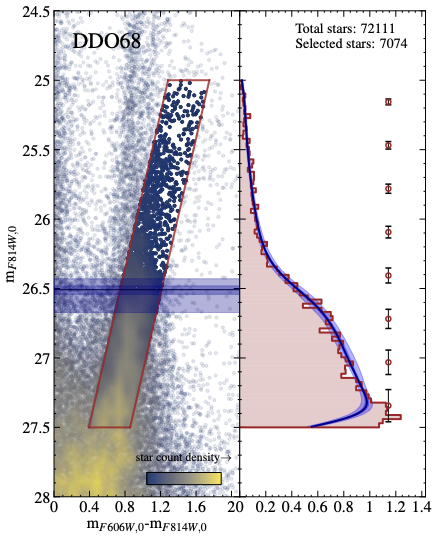}}
{\includegraphics[width=0.45\textwidth]{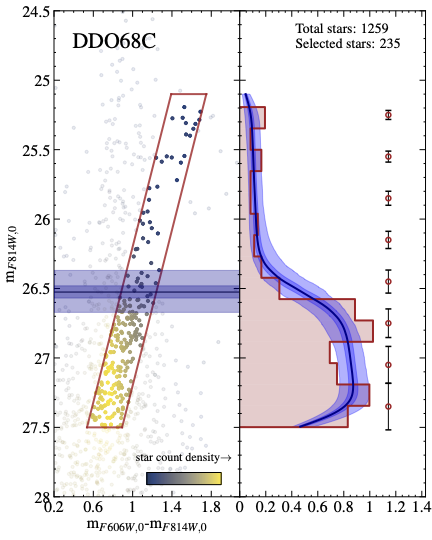}}}
\caption{Detection of the TRGB in DDO~68 (left pair of panels) and in DDO~68-C (right pair of panels). In each pair of panels the CMD (zoomed on the RGB region) is shown as a Hess diagram and is flanked by the LF of the stars selected for the detection of the tip, following BP24. In the CMDs the adopted selection box is drawn in red and the selected stars highlighted; the TRGB level is marked by a black horizontal line while the dark and pale purple stripes display the $\pm 1\sigma$ and $\pm 3\sigma$ uncertainty intervals, respectively. In the flanking panels the pink histogram is the observed LF, the black line is the best-fit model, the dark and pale purple stripes display the $\pm 1\sigma$ and $\pm 3\sigma$ uncertainty intervals, and the error bars are the photometric uncertainties as a function of magnitude.}
\label{fig:trgb} 
\end{figure*}

\section{DDO~68-C distance from the TRGB} 
\label{sec:trgb}

To test the hypothesis that DDO~68-C is physically associated with DDO-68, we need to verify that the distances of the two systems are compatible. To do so, we use the TRGB \citep[][see also BP24 and references therein]{mf95,sc98} as the best standard candle that is available in our datasets. In particular, we used the approach described in detail in BP24 that allows us to obtain a TRGB estimate by fitting a model to the RGB luminosity function (LF) within a fully Bayesian framework \citep[following, e.g.,][]{Makarov2006,Conn2011}. The actual measures, performed on reddening-corrected magnitudes in the native ACS/WFC photometric system, are illustrated in Fig.~\ref{fig:trgb}, where the selection of the sample of RGB+AGB stars is depicted, as well as their LF and the corresponding best-fit model. The results are summarized in Table~\ref{tab:tab1}. 



\begin{deluxetable}{cccc}
\tablecaption{Magnitude and color of the TRGB \label{tab:tab1}}
\tablehead{\colhead{ID} & \colhead{ m$_{F606W,0}-$m$_{F814W,0}$} & \colhead{m$_{F814W,0}$ } & \colhead{$\pm1(3)\sigma$}} 
\startdata
DDO~68 &     0.930$^{+0.137}_{-0.118}$ &   26.509 &    $^{+0.034(+0.164)}_{-0.027(-0.079)}$ \\
DDO~68-C &   1.078$^{+0.074}_{-0.124}$ &   26.525 &    $^{+0.043(+0.146)}_{-0.045(-0.156)}$  \\
\enddata
\tablecomments{The color is the median color of the stars within a $\pm 1\sigma$ strip around the TRGB; the associated
uncertainties are the differences between the mean and the 16th and 84th percentile of the color distribution within the same strip, following BP24.}
\end{deluxetable}


As can be visually appreciated from Fig.~\ref{fig:trgb}, the RGB of DDO~68-C is sparsely populated and therefore the measurement of the TRGB is uncertain. In this context, we verified that small changes on the choice of the blue edge of the selected RGB region, where the possible contamination from young stars can bias the results given the low number statistics, do not affect significantly the measured TRGB magnitude (i.e., within $\pm$ 0.02 mag). 
However, it is evident at a first glance that the two galaxies should lie at very similar distances. Indeed, the m$_{F814W,0}$ magnitudes of their TRGB are fully compatible with this hypothesis, as they differ by only $0.016\pm 0.056$~mag, while the respective mean TRGB colours are similar, within the uncertainties.

Adopting $M_{F814W_{0}}^{TRGB}$= -3.996$\pm$ 0.045 from BP24, appropriate for the TRGB colors of both galaxies, we obtain  a distance modulus $(m-M)_{0}=30.505\pm 0.056$ for DDO~68, corresponding to a distance D$=12.6 \pm 0.3$~ Mpc, in good agreement with the results in the literature \citep[see, e.g.,][]{Cannon2014,Sacchi2016}, whereas for DDO~68-C we obtain $(m-M)_{0}=30.521\pm 0.064$, corresponding to a distance D$=12.7 \pm 0.4$~Mpc. The final uncertainties on the distance modulus are obtained by adding in quadrature the uncertainty on the measure of the TRGB obtained here with the uncertainty on the calibrating relation reported in BP24. Taking these results at face value, we find that the difference in distance is just $\Delta D =0.1\pm 0.5$~Mpc, less than 0.2$\sigma$, with DDO~68-C appearing slightly more distant than DDO~68. 
   
Hence, when all possible sources of uncertainty are properly taken into account, we can conclude that the obtained results are fully compatible with the hypothesis that the two galaxies are very close to each other. It is important to note that the degree of homogeneity of the analysis is very high: the two measures that we compared are derived from photometric observations obtained with the same instrument and same passbands, reduced with the same state-of-the-art photometric code, with samples selected in the same way and reddening corrections from the same sources, adopting the same reddening laws (both galaxies being affected by low extinction, $E(B-V)<0.02$). The detection of the TRGB has equally been performed with the same method and converted to distance with the same calibration. 

\section{DDO~68-C structural parameters}
\label{sec:structure}

\begin{figure*}[thbp!]
\begin{center}
{\includegraphics[width=0.9\textwidth]{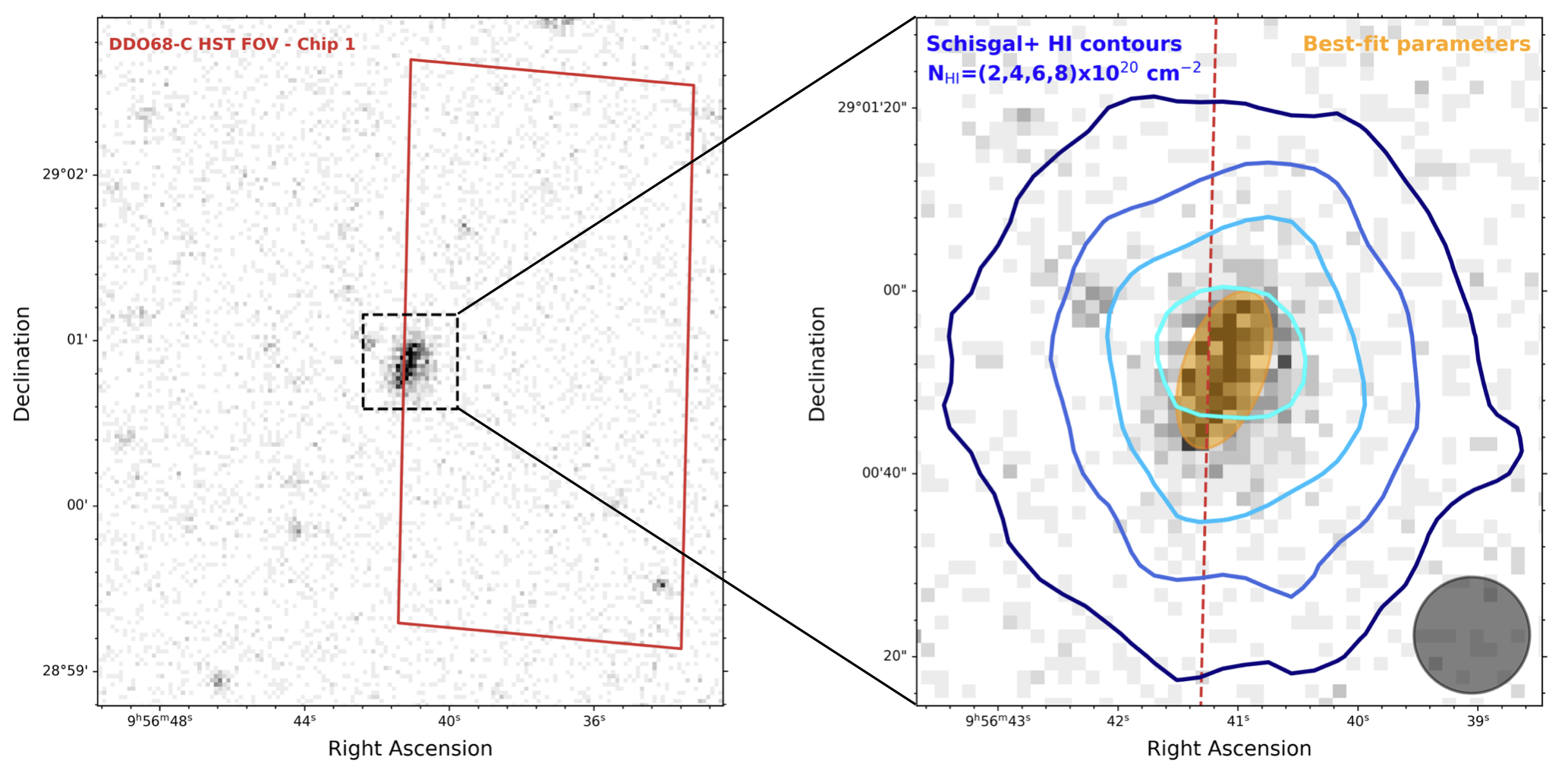}}
\end{center}
\caption{Left panel: GALEX FUV image with overplotted in red the field of view of the HST chip 1 showing the fraction of DDO~68-C covered by our observations. Right panel: zoom-in of the FUV image in the DDO~68-C region; the best-fit parameters (i.e., center coordinates, $R_e$, ellipticity, and position angle) derived from our analysis are represented by the orange ellipse, whereas with the different shades of blue we depict the H~I contours obtained from new VLA observations (Schisgal et al., in preparation) adopting a synthetised beam size of $\sim$ 13 arcsec (depicted in the bottom right corner). The red dashed line indicates the edge of the HST field of view.}
\label{fig:structural_parameters} 
\end{figure*}

The measurement of parameters as fundamental as the integrated luminosity and the effective radius is a challenging task in the case of DDO~68-C, since, even if we were able to resolve stars over a significant fraction of the galaxy with both HST (see left panel of Fig.~\ref{fig:structural_parameters}) and LUCI-SOUL \citep{Annibali2023} our optical and NIR observations do not cover the entire body of the galaxy. The bright star intervening along the line of sight prevents a standard analysis of the overall light-distribution and we had to adopt sub-optimal procedures. However, these measures, albeit uncertain, are useful to put DDO~68-C in the proper context as a dwarf galaxy.

We made various attempts to estimate the parameters of the light profile from our photometric catalogue of resolved stars, e.g., by modifying the method by \citet{Martin2008} to account for the missing portion of the galaxy, but we were not able to find satisfactory solutions.
We concluded that the only viable way to constrain the light profile of DDO~68-C is by fitting the only image where the entire body of the galaxy is visible, i.e. the GALEX FUV image. Hence, we performed a 2D fit of this image using {\tt imfit} \citep{Erwin2015}. {\tt imfit} allows the generation of a model image, using one or more 2D models of the light distribution (e.g., exponential, S{\'e}rsic, etc.), that is matched to the input image with parameter optimization obtained through nonlinear minimisation of the total $\chi^2$. 

The best-fit to the FUV image of the galaxy has been achieved adopting a \citet{sersic} profile with S{\'e}rsic index {\it n}=0.68$^{+0.19}_{-0.09}$, and effective radius $R_e=9.0 \pm 1.0$~arcsec along the major axis. The circularised value is $R_{e,circ}=R_e\sqrt{1-e}=6.4\pm 0.7$~arcsec, corresponding to $394\pm 43$ pc at the distance of DDO~68-C. The position angle is {\it PA}=160.8$^{+2.6}_{-2.1}$ deg, from N toward E, and the ellipticity {\it e}=0.50$^{+0.03}_{-0.04}$, compatible with the distribution of stars in the ACS optical images. The uncertainties on the derived parameters have been computed by the boot-strap procedure embedded in {\tt imfit}.
The coordinates of the center of the best-fit model are in excellent agreement with those derived by \cite{Cannon2014}. 

The right panel of Fig.~\ref{fig:structural_parameters} shows the FUV image, zoomed-in on DDO~68-C, with overplotted an ellipse with ellipticity, and position angle of the best-fit model and major axis corresponding to the effective radius $R_e$, and H~I contours derived from from new observations acquired with the National Science Foundation's Karl G. Jansky Very Large Array (VLA\footnote{The National Radio Astronomy Observatory and Green Bank Observatory are facilities of the U.S. National Science Foundation operated under cooperative agreement by Associated Universities, Inc.}) in programs 23A-195 and 23B-131  (Schisgal et al., in preparation) which confirm the co-location of the stellar and gas emission. 
S{\'e}rsic indices $n<1.0$ are typical of dwarf galaxies fainter than $M_B\simeq -15.0$ \citep{Cote2008} and the effective radius lies in the range spanned by galaxies with  $-15.0 \la  M_V \la -10.0$ \citep{mc12}.

\begin{deluxetable}{lccc}
\tablecaption{Properties of DDO~68-C \label{tab:tab2}} 
%
\tablehead{\colhead{Parameter } & \colhead{Value } & \colhead{Unit} & \colhead{Ref.}}  
\startdata
ra$_{ICRS}$  &  149.171312  &  deg  &  t.w.\tablenotemark{a} \\
dec$_{ICRS}$  &  +29.0142707  &  deg  &  t.w.  \\
D  &  $12.7\pm 0.4$   &  Mpc  &  t.w. \\
E(B-V)  &  0.016  &  mag  &  SF11\tablenotemark{b} \\
$M_V$  &  $-13.1\pm0.5$  &  mag  &  t.w. \\
ell  &  $0.50^{+0.03}_{-0.04}$  &  \nodata  &  t.w. \\
PA  &  $160.8^{+2.6}_{-2.1}$  &  deg  &  t.w.  \\
R$_e$   &  $9.0\pm 1.0$  &  arcsec  &  t.w.\tablenotemark{c} \\
R$_{e,circ}$   &  $6.3\pm 0.7$  &  arcsec  &  t.w. \\
R$_{e,circ}$   &  $394\pm 43$  &  pc  &  t.w. \\
M$_{\star}$  &  $\simeq 1.5\times10^7$  &  M$_{\sun}$  &  t.w.  \\
M$_{HI}$  &  $(2.8\pm 0.5)\times 10^7$  &  M$_{\sun}$   &  C14\tablenotemark{d} \\
$V_{sys}$  &  $505.5\pm 1.0$  &  km~s$^{-1}$  &  C14\tablenotemark{d} \\
\enddata
\tablenotetext{a}{This work}
\tablenotetext{b}{Reddening values are from the \citet{sfd98} maps, re-calibrated according to \citet[][SF11]{sf11}.}
\tablenotetext{c}{measured along the major axis.}
\tablenotetext{d}{\citet[][C14]{Cannon2014}.}
\end{deluxetable}

On the other hand, to get constraints on the integrated magnitude we used simple aperture photometry on our ACS/WFC images. Using the IRAF task {\it polyphot} we measured the total flux enclosed inside a polygonal aperture enclosing all the stars of DDO~68-C identified in those images, after masking the spikes from TYC~1967-1114-1 and other contaminants that fall inside the polygon (e.g., background galaxies).
We used {\it polyphot} also to estimate the background, selecting a wide region just outside the aperture adopted for the galaxy, masking obvious contaminants as above. Transforming from fluxes to VEGAMAG magnitudes 
we obtained m$_{F606W} = 17.33$ and m$_{F814W} = 16.60$ mag. The uncertainty on these values is dominated by two factors: on one side by the fact that we are missing part of the light of the galaxy as it lies beyond the eastern limit of our FoV, on the other side by the fact that our aperture photometry includes some contribution from the wings of TYC~1967-1114-1. The first factor has clearly the largest amplitude of the two. From our images (i.e., Fig.~\ref{fig:fov} and Fig.~\ref{fig:structural_parameters}) and from the structural parameters derived above we know that our ACS images miss less than half of the main body of the galaxy, hence the total light should be underestimated by a factor smaller than two, that is less than 0.75 mag. In the following we assume, conservatively, an uncertainty on the integrated magnitudes of $\pm 0.5$~mag.
Adopting Eq.~1 of  \cite{Galleti2006} to transform our F606W, F814W integrated magnitudes into the corresponding Johnson-Kron-Cousins V magnitude, and the appropriate  distance modulus and reddening, we obtain $M_V =-13.1 \pm 0.5$, within the range of compatibility with the absolute magnitude - $R_h$ relation quoted above, and implying M$_{\star}\simeq 1.5\times10^7$M$_{\sun}$, adopting M/L$_V=1.0$, for simplicity. The most relevant properties of DDO~68-C are summarized in Table~\ref{tab:tab2}.

\section{Discussion}
\label{sec:discussion}

While a physical association between DDO~68 and DDO~68-C was initially suggested by \cite{Cannon2014} based on their small projected distance of $\simeq$43 kpc, very similar radial velocities, and presence of a low surface brightness H~I bridge connecting the two, our analysis places for the first time this hypothesis on a secure footing by demonstrating that the two galaxies have also very similar line-of-sight distances.  
Indeed, through an homogeneous analysis of the HST photometry of both systems, we infer TRGB distances of 
$=12.6 \pm 0.3$~Mpc and $=12.7 \pm 0.4$~Mpc for DDO~68 and DDO~68-C, respectively, providing a line-of-sight distance difference of just $\Delta D =0.1\pm 0.5$~Mpc.

The typically low density of galaxies in void environments makes the presence of two close systems like DDO~68 and DDO~68-C an uncommon configuration. In the Lynx-Cancer Void, which is centered at $\sim$18 Mpc distance from us and extends for a radius of $\sim$8.2 Mpc, 75 dwarf galaxies with $-11.9<M_B<-18$ were identified by \cite{Pustilnik2011}. From the sample properties listed in Table 2 of their paper, it results that only $\sim$13\% of these galaxies have a neighbour at a 3d-distance $\lesssim$0.1 Mpc, with the median minimum distance between galaxies being $\sim$1 Mpc. 
Indeed, \cite{Pustilnik2011} noticed that  
a non-negligible fraction of the Lynx-Cancer Void galaxies are paired with typical projected distances of several tens of kpc. According to our results, the DDO~68 - DDO~68~C system fits into this population of paired dwarfs, but it is also exceptional in that the former member also hosts two disrupted satellites,
DDO~68-B \citep{Tikhonov2014} and S1 (A19), a configuration that to our knowledge has not been observed so far for other members of the Lynx-Cancer Void. 
Interestingly, \cite{Pustilnik2011} also noticed the presence of several unbound, $\sim$1-2 Mpc size elongated structures and identified DDO~68, together with UGC~5427 and KISSB~23, to be part of such a ``filaments''.  This configuration is reminiscent of other associations of dwarf galaxies that are organized along elongated structures, such as the  NGC~3109  association \citep{Bellazzini2013} and the isolated group of star-forming dwarfs recently identified by \cite{Paudel2024}. Whether DDO~68 and its two satellites,  the just confirmed DDO~68-C neighbour, and the two more distant dwarfs UGC~5427 and KISSB~23 are all part of a common filament is a tantalizing hypothesis that nevertheless can not be confirmed with the data in hand (see \citealt{Tully06} for a more general view on groups of dwarf galaxies). 

Accretion of metal-poor gas either from the intergalactic medium or from a metal poorer companion was proposed as a viable scenario in order to explain DDO~68's extremely low metal content and its large deviation from the (L-Z) and (M-Z) relations.  However, \cite{Pascale2022} showed that the gas contribution provided by satellite B, ten times less massive than DDO~68, would have been insufficient to accomplish a full dilution from a metallicity as high as 12 + log(O/H) $\simeq$7.9 (the value expected from the M-Z relation in \cite{Berg2012} and DDO~68's stellar mass of M$_{\star}\simeq 10^8$M$_{\odot}$) to the observed H~II region value of  12+log(O/H) $\leq$ 7.2, even assuming a zero-metallicity gas for DDO~68-B. The same considerations may apply to DDO~68-C, whose stellar and gaseous masses of M$_{\star} \simeq 1.5\times 10^7$ M$_{\odot}$ and M$_{HI} \simeq 2.8\times 10^7$ M$_{\odot}$, respectively, (see Table~\ref{tab:tab2}), appear incompatible with the hypothesis that this dwarf has been responsible for diluting a total DDO~68's gas mass of M$_{HI}\simeq10^9$ M$_{\odot}$ \citep{Cannon2014} down to a $\simeq$3\% Solar metallicity.

Indeed, this would require the original gas content of DDO~68-C to be at least as high as $\sim$75\% of the total gas mass of DDO~68 \citep[see section~5.2 of][]{Pascale2022}, yielding an unrealistic gas-to-mass ratio of $\sim$50. 

In this context, \cite{Pascale2022} suggested that the large offset in the M-Z relation for DDO~68 can be explained as a result of a misclassification, since the H~II regions may be tracing the metallicity of its $10^7$ M$_{\odot}$ accreted companion and not that of the main body at $10^8$ M$_{\odot}$. 

While the detection by \cite{Cannon2014} of a low surface brightness H~I bridge connecting DDO~68 and DDO~68-C suggests 
an interaction between the two systems, their relative distance of at least $\simeq$43 kpc coupled with typical peculiar velocities of $\simeq$50-100 km/s for galaxies in void environments \citep[e.g.][]{Pustilnik2011}
seem to exclude the possibility of a close encounter over the last $\sim$400 Myr. 
Nevertheless, even in the scenario of a first infall between the two systems, it is still possible to imagine some  effect on the star formation history of DDO~68-C from its more massive companion. 
Indeed, from the study of a large sample of dwarf galaxies drawn from the SDSS, \cite{Stierwalt2015} revealed 
that paired dwarfs present an enhancement in their star formation rate compared to otherwise isolated systems. Interestingly, they also noticed that the enhancement, which increases with decreasing pair distances, is present 
as far as to separations of  $\simeq$100 kpc, corresponding to roughly a virial radius for 
the typical galaxy mass of M$_{\star}\simeq10^9 $M$_{\sun}$ of their sample. Assuming that their results can be extended to the lower mass regime of DDO~68 and adopting the virial radius - mass scaling relation of \cite{Kravtsov2013}, 
we may expect some disturbance on the star formation of a companion as far as $\simeq$50 kpc, a situation  compatible with the DDO~68 - DDO~68-C distance.

We discussed in Section~\ref{sec:cmd} hints for the presence of an enhanced star formation activity event in DDO~68-C about $\simeq$100 Myr ago; more quantitative results based on synthetic CMD modeling, that will be presented in a future paper devoted to a larger sample of dwarf galaxies with close companions, will be crucial to confirm the occurrence of starburst events potentially triggered by the interaction with DDO~68. 
Given the ten times higher mass of DDO~68 compared to DDO~68-C, it is reasonable to expect that the gravitational interaction between the two systems will have negligible effects on the star formation of the former compared to those  induced on its lower mass companion. 

\section{Summary and conclusions}
\label{sec:summary}

We acquired new deep HST ACS data of the dwarf galaxy DDO~68-C and, for the first time ever, 
were able to resolve its stellar content over a significant fraction of the galaxy down to $\simeq$1.5 mag below the TRGB.  Our study shows that: 

\begin{itemize}

\item The distance to DDO~68-C is $=12.7 \pm 0.4$~Mpc, very similar to the distance of $=12.6 \pm 0.3$~Mpc inferred for DDO~68 through a homogeneous re-analysis of the archival HST data. 
This provides a nominal line-of-sight distance difference of just $\Delta D =0.1\pm 0.5$~Mpc and demonstrates the physical association between the two systems, separated by a projected distance of only $\sim$43 kpc.

\item The resolved star CMD exhibits all the populations typically present in star forming galaxies, from 
very young (age$\lesssim$10-20 Myr) up to several Gyrs old. There are hints for an enhanced  star formation event 
around $\simeq$100 Myr ago, to be confirmed through proper synthetic CMD modeling. Both the old and the young stars are compatible with a few percent Solar metallicity, indicating a very modest chemical enrichment with time. 

\item From the new HST data, we infer for DDO~68-C an absolute magnitude of $M_V =-13.1 \pm 0.5$ and 
a stellar mass of $M_{\star}\simeq 1.5\times10^7~M_{\sun}$. From a two dimensional fit of the GALEX FUV images, we find that the galaxy is well described by a \citet{sersic} profile with S{\'e}rsic index {\it n}=0.68$^{+0.19}_{-0.09}$ and effective circularised radius of $R_{e,circ}=6.3\pm 0.7$~arcsec or $382\pm 46$ pc. 
These parameters are in agreement with typical properties and scaling relations of dwarf galaxies with 
 $M_V\gtrsim-15$.

\item Like the DDO~68 - DDO~68-C system, other paired dwarfs with physical separation $\lesssim$0.1 Mpc are found in the Lynx-Cancer Void, with an occurrence of about $\simeq$13\% of the total dwarf sample.  
Nevertheless, the DDO~68 - DDO~68-C pair is an exceptional system in that DDO~68 also hosts two disrupted satellites (the DDO~68-B component and the S1 stream), a configuration that has no counterpart in the  Void. 

\item While the very presence of a low surface brightness H~I bridge connecting DDO~68 and DDO~68-C suggests 
a gravitational interaction between the two systems, we can exclude that a close encounter has occurred within the last $\sim$400 Myr, given the separation between the two dwarfs and the typical relative velocities of galaxies in voids. It is also very unlikely that DDO~68-C has been responsible for the extremely low gas metallicity in DDO~68. Still, we can not exclude that the gravitational interaction between the two systems has produced some disturbance on the star formation history of DDO~68-C.

\end{itemize}

%

\section*{Acknowledgments}
We warmly thank R. van der Marel for his competent support.

M. Correnti acknowledges financial support from the ASI-INAF agreement n.2022-14-HH-0. These data are associated with the HST GO program 17131 (PI: F. Annibali). Support for program number 17131 was provided by NASA through a grant from the Space Telescope Science Institute, which is operated by the Association of Universities for Research in Astronomy under NASA contract. 
F. Annibali and M. Bellazzini acknowledge financial support by INAF, grant Ob. Fu. 1.05.23.05.09 "Dwarf galaxies as probes of the Lambda Cold Dark Matter hierarchical paradigm at the smallest scales" (P.I.: F. Annibali). R. Pascale acknowledges support by the Italian Research Center on High Performance Computing Big Data and Quantum Computing (ICSC), project funded by European Union - NextGenerationEU - and National Recovery and Resilience Plan (NRRP) - Mission 4 Component 2 within the activities of Spoke 3 (Astrophysics and Cosmos Observations).

\section*{Data Availability}
MAST data underlying this article are available at doi:
http://dx.doi.org/10.17909/s2rd-4p19

%

\vspace{5mm}
\facilities{HST(ACS/WFC), GALEX(FUV), VLA}


\software{{\tt DOLPHOT} \citep[\url{http://americano.dolphinsim.com/dolphot/};][]{Dolphin2000, Dolphin2016}, {\tt imfit} \citep[\url{https://www.mpe.mpg.de/~erwin/code/imfit/};][]{Erwin2015}, {\tt Astropy} \citep{Astropy2013}, {\tt Numpy} \citep{Numpy2011}, {\tt Pandas} \citep{Pandas2010}, {\tt SciPy} \citep{Scipy2020}, {\tt Matplotlib} \citep{Hunter2007}, {\tt APLpy} \citep{Robitaille2012, Robitaille2019}
}




    



\end{document}